\documentclass[twocolumn,amsmath,amssymb,floatfix,superscriptaddress,prb,longbibliography]{revtex4-2}
\usepackage{lineno,hyperref}
\usepackage{bm}
\usepackage{amsmath}
\usepackage{gensymb}
\usepackage{epstopdf}
\usepackage{booktabs}
\usepackage{longtable}
\usepackage{tabularx}
\usepackage{multirow}
\usepackage{setspace}
\usepackage{mathrsfs}
\usepackage{tabularray}

\usepackage{graphicx}
\usepackage{ulem}
\usepackage{xcolor}
\usepackage{hyperref}
\usepackage{amsfonts} 
\usepackage{siunitx}
\usepackage[version=4]{mhchem}

\begin{document}

\preprint{APS/123-QED}

\title{Disproportionate influence of site disorder on the evolution of magnetic phases in anti-Heusler alloy Al$_2$MnFe}
\author{Soumya Bhowmik}
\affiliation{Condensed Matter Physics Division, Saha Institute of Nuclear Physics,
              A CI of Homi Bhabha National Institute, 1/AF Bidhannagar, Kolkata 700064, India}
\author{Santanu Pakhira}
\affiliation{Department of Physics, Maulana Azad National Institute of Technology, Bhopal, M.P. - 462003, India}
\author{Ashis Kundu}
\affiliation{Department of Physics, Indian Institute of Science Education and Research (IISER) Pune, P.O. 411008, India}
\author{V. Raghavendra Reddy}
\affiliation{UGC-DAE Consortium for Scientific Research, Khandwa Road, Indore 452001, India}
\author{Mukul Kabir}
\affiliation{Department of Physics, Indian Institute of Science Education and Research (IISER) Pune, P.O. 411008, India}
\author{Chandan Mazumdar}
\email{chandan.mazumdar@saha.ac.in}
\affiliation{Condensed Matter Physics Division, Saha Institute of Nuclear Physics,
              A CI of Homi Bhabha National Institute, 1/AF Bidhannagar, Kolkata 700064, India}

\date{\today}

\begin{abstract}

Anti-Heusler alloys, being a new addition to the Heusler alloys family, exhibit atomic disorders, and almost all of them are reported as a re-entrant spin-glass system. Although such spin-glass feature is generally attributed to the inherent atomic disorder, a comprehensive and extensive investigation on the individual roles of different types of disorders in magnetic interactions remains lacking for any of the reported anti-Heusler systems. As an illustrative case, we have carried out an in-depth experimental as well as theoretical investigation of structural, magnetic, and transport properties of a polycrystalline anti-Heusler alloy, Al$_2$MnFe. While the major atomic disorder is found to be among Fe and Mn atoms, which are randomly distributed among the two octahedral sites, 4$a$ and 4$b$ (B2-type disorder), a relatively small fraction ($\sim$12\%) of Mn atoms also replace Al atoms at the tetrahedral 8$c$ site. Magnetically, the system undergoes two transitions: a paramagnetic to a ferromagnetic transition at $T_{\rm C}\sim$113~K, followed by a spin-glass phase transition below $T_{\rm f}\sim$20~K. Here, the magnetic moment is primarily confined to Mn atoms. Very interestingly, our theoretical analysis reveals that the ferromagnetic spin arrangement remains rather robust in spite of the 50\% disorder of moment-carrying Mn atoms between the two octahedral sites, but a much smaller ($\sim$12\%) cross-distribution of Mn atoms between octahedral and tetrahedral sites are sufficient to impose a reentrant spin-glass state at low temperature. Our analysis brings forth the importance of understanding the role of individual types of swap-disorder on magnetic properties in the anti-Heusler family of materials.

\end{abstract}

\maketitle

\section{Introduction}
Structural disorder, a natural trait of most of the real materials, plays a crucial role in determining various physical properties such as thermal conductivity~\cite{Zhang2021,Garmroudi2022, Chiritescu2007, Simonov2020}, electrical transport~\cite{PhysRevLett.87.186807,PhysRevB.107.045407}, and magnetization~\cite{PRXEnergy.3.033006,PhysRevB.98.094418,PhysRevB.62.3296, PhysRevB.107.014108}. In the pursuit of advancements in spintronics, where phenomena are governed by spin degrees of freedom, understanding the interplay between structural disorder and spin degrees of freedom has become essential. This is particularly important for Mn-based Heusler alloys (HA), where Mn can exhibit both itinerant and local-moment behavior. The presence of atomic disorder in these materials can introduce competing magnetic interactions and magnetic frustration, leading to complex spin arrangements. As a result, such systems often exhibit non-collinear spin structures~\cite{Singh2016,PhysRevLett.113.087203} and other exotic magnetic ground states~\cite{Nayak2017}, including spin-glass~\cite{PhysRevB.99.174410,PhysRevB.107.184408} and re-entrant spin-glass phases~\cite{PhysRevB.79.092410, PhysRevB.108.054405,10.1063/1.3651767}.

Due to the similar electronic properties and atomic radii of their constituent elements, HA inherently exhibit antisite and (or,) swap atomic disorders among the constituent elements, leading to a wide range of intriguing physical phenomena~\cite{GRAF20111}. These properties are half-metallic ferromagnetism~\cite{PhysRevLett.50.2024,Jourdan2014,PhysRevB.106.115148,PhysRevB.108.045137}, spin-gapless semiconductivity~\cite{PhysRevLett.110.100401,PhysRevB.91.104408,PhysRevB.97.054407}, the giant magnetocaloric effect~\cite{Liu2012}, thermoelectricity~\cite{PhysRevB.98.205130,10.1063/5.0043552,adfm.201705845}, and complex non-collinear magnetic orderings such as magnetic skyrmions~\cite{Nayak2017}, topological properties, spin-glass behavior~\cite{PhysRevB.99.174410,PhysRevB.107.184408}, and Weyl semimetallicity~\cite{Manna2018}. Due to the high chemical tunability, the physical properties of HA can be easily modified through doping, elemental substitution, or even structural alterations. This flexibility has led to the exploration of various types of HA, including full ($X_2YZ$), inverse ($Y_2XZ$), half ($XYZ$), and quaternary ($XX^{\prime}YZ$) Heuslers, where $X, X^{\prime}$ and $Y$ is the transition element and $Z$ is the $p$-block element~\cite{GRAF20111}.

Recently, a new kind of HA has been formed by substituting the $Z$ element by $X$ and vice versa in a full HA ($X_2YZ$) structure (Fig.~\ref{Fig:Anti-Heusler structure}). In contrast to the known HAs, it comprises half $p$-block and half transition-elements with a reversing formula unit, $Z_2XY$~\cite{Samanta2022}, thus named as anti-HA~\cite{PhysRevB.111.174417}. The enhancement of $p$-block element concentration (50\%) in this anti-HA may influence the band structure and manifest different physical as well as magnetic properties from usual full-, inverse-, quaternary- (25\%), and half-(33\%) HAs.  Only a few of these alloys have been studied so far, and interestingly, almost all of those exhibit reentrant spin glass behavior, where the system first undergoes long-range magnetic ordering at a higher temperature before transitioning into a glassy state at lower temperatures. While some efforts have been made to explain the magnetic ground state, a significant research vacuum remains in understanding the fundamental origins of the development of spin-glass state within the already ordered ferromagnetic state in this class of materials~\cite{PhysRevB.78.134406,PhysRevB.97.184421,Samanta2022}.

In the anti-HA family, Al$_2$MnCu is the only reported compound that exhibits long-range ferromagnetic ordering without transitioning into a spin-glass state~\cite{PhysRevB.111.174417}. However, substituting Cu with Co in Al$_2$MnCo leads to the emergence of a spin-glass state below the ferromagnetic ordering temperature~\cite{SHIRAISHI19922040}. Since Co has an electronic structure and atomic radius more similar to Mn in comparison to those of Cu, the likelihood of antisite and swap disorder is expected to be higher in Al$_2$MnCo than in Al$_2$MnCu. This increased disorder may contributes to the observed additional spin-freezing behavior at low temperatures. To further investigate the origin of atomic disorder and its role in spin-glass formation within the anti-HA family, we examine Al$_2$MnFe, where an even higher degree of antisite and swap disorder is anticipated due to the close proximity of Mn and Fe in the periodic table.

In this work, we have synthesized polycrystalline Al$_2$MnFe and carried out a comprehensive experimental study of its structural, magnetic, and transport properties, that was also complemented by theoretical DFT calculations. The primary objective is to identify the nature and extent of atomic disorder within the system and elucidate its role in shaping the observed magnetic and magnetotransport behaviors. A correlation between the intrinsic disorder developed in the system and the observed magnetic and associated magnetotransport anomalies is discussed to have deeper insight into the magnetism in the anti-HA class of materials on the verge of structural and elemental diversities. Such an understanding of magnetism and its intertwined transport properties is essential for the practical realization and development of novel anti-HA for spintronics applications.

\section{Methods}
\subsection{Experimental details}
The compound, Al$_2$MnFe was prepared in poly-crystalline form by employing tri-arc furnace, where the pure constituent elements ($>$99.9 wt.\%, M/s Alfa Aesar, USA) were melted in an Argon gas atmosphere. For better homogeneity, the sample was remelted several times and flipped after each melting process. Apart from the stoichiometric elemental ratio, extra 2\% weight of Mn was added for taking care of its evaporation during the melting process. To ensure its single phase nature and to obtain the information of crystal structure, powder X-ray diffraction (XRD) was carried out at room-temperature by employing a diffractometer (rotating Cu anode, 18 kW, Model: TTRAX-III, M/s Rigaku Corp., Japan). The XRD spectra was analysed using the FULLPROF software package~\cite{RODRIGUEZCARVAJAL199355}, where the crystal structure was determined by performing Rietveld refinement method.

The dc magnetization measurement was performed using commercial super-conducting quantum interference device (SQUID) magnetometer in the magnetic field regime of 0-70 kOe and the temperature range of 2–400~K (Quantum Design, Inc., USA). The temperature dependent magnetization data is taken in zero-field-cooled (ZFC) and field-cooled (FC) protocols. In ZFC protocol, the sample was cooled down from its paramagnetic state without applying any field, whereas the same was done with field in FC method. Subsequently, the magnetization of the sample was investigated with the same externally applied field, under which the sample was cooled (FC), otherwise, a field was applied (ZFC), and the magnetization data was recorded during the heating cycle. Beside that, the field dependent isothermal magnetization data was collected by stabilizing at each temperature, after cooling down from the paramagnetic region of the sample before each such isothermal measurements.

The ac susceptibility measurement was carried out in the temperature range of 2-175~K with different excitation fields ranging between 0.5-10~Oe and at various frequencies over a span of 113 to 9951~Hz using a
physical property measurement system (M/s Quantum design, Inc., USA) with ACMS attachment. For both the frequency as well as excitation field dependent measurements were done in heating cycle, each time after cooling from the paramagnetic state of the sample. The heat capacity of the studied compound was measured by standard relaxation method in the same PPMS with heat capacity attachment. DC transport measurement has been carried out using the standard four-probe method on a rectangular-shaped sample having approximate dimensions of $3~{\rm mm} \times 1.5~{\rm mm} \times0.5~{\rm mm}$. The transverse Hall and longitudinal magneto-resistance data were recorded in the temperature range of 5-300~K, with applied magnetic field variation of -70~kOe to +70~kOe. The Hall resistivity data was acquired by anti-symmetrizing the raw data with respect to the externally applied magnetic field whereas the longitudinal magneto-resistance data were symmetrized to separate out of each other contribution. $^{57}$Fe Mössbauer measurements were carried out in transmission mode with $^{57}$Co radioactive source in constant acceleration mode using standard PC-based Mössbauer spectrometer equipped with WissEl velocity drive. Velocity calibration of the spectrometer is done with natural iron absorber at room temperature. Low temperature measurements are carried out using Janis make superconducting magnet. The spectra were analyzed with NORMOS program.

\subsection{Computational details}
Electronic structure calculations were performed using spin-polarized density functional theory (DFT) based on the projector augmented-wave (PAW) method~\cite{BlochPRB94}, as implemented in the Vienna ab initio simulation package (VASP)\cite{KressePRB93,KresseCMS96,KressePRB96}. For all calculations, we used the Perdew–Burke–Ernzerhof (PBE) implementation of the generalized gradient approximation (GGA) for the exchange-correlation functional\cite{PerdewPRL96}. An energy cutoff of 500 eV was employed. The convergence criteria for the total energies and atomic forces were set to $10^{-6}$ eV and $10^{-2}$ eV/\AA, respectively. In order to model the disordered structure, we used a 128-atom Special Quasi-random Structure (SQS)\cite{ZungerPRL90}. The SQS was generated using a $2\times2\times2$ supercell of the conventional unit cell (16 atoms) of a Heusler system with the ATAT package\cite{WalleCAL13}. For example, in the case of B2-type disorder in Al$_2$MnFe, Fe atoms at the $4a$ site (32 atoms) and Mn atoms at the $4b$ site (32 atoms) were equally intermixed within the supercell. Based on careful testing, a cutoff of 8~\AA\ for pairs and 6~\AA\ for triplets was used to generate a well-randomized SQS. For the comparison of total energies, we also used a 128-atom supercell for the ordered structure. All calculations were performed using a Monkhorst–Pack~\cite{MethfesselPRB89} $2\times2\times2$ k-point mesh for structural optimization and a $3\times3\times3$ mesh for self-consistent total energy calculations. The experimentally measured lattice constant was used in all cases.


The magnetic pair exchange parameters are calculated to gain insight into the magnetic interactions within the system. This calculation is done using the multiple scattering Green function method implemented in the SPRKKR package~\cite{EbertRPP11}. In this approach, the spin component of the Hamiltonian is mapped to the Heisenberg model.
\begin{eqnarray}
H = -\sum_{\mu,\nu}\sum_{i,j}
J^{\mu\nu}_{ij}
\mathbf{e}^{\mu}_{i}
.\mathbf{e}^{\nu}_{j}
\end{eqnarray}
$\mu$, $\nu$ represent different sub-lattices, $i$, $j$ represent atomic positions and $\mathbf{e}^{\mu}_{i}$ denotes the unit vector along the direction of magnetic moments at site $i$ belonging to sub-lattice $\mu$. The $J^{\mu \nu}_{ij}$ are calculated from the energy differences due to infinitesimally small orientations of a pair of spins within the formulation of Liechtenstein \textit{et al.}~\cite{LiechtensteinJMMM87}.  We use a full potential spin-polarized scalar relativistic Hamiltonian with an angular momentum cutoff of $l_{max} = 3$ and a converged $k$-mesh for Brillouin zone integration. The Green's functions were computed for 56 complex energy points arranged along a semi-circular contour. A convergence criterion of 10$^{-6}$ Ry was set for the self-consistency cycles. The calculations used the lattice constant obtained from our experimental measurements.

\section{Experimental Results}

\subsection{Structural details}

\begin{figure} [h]
  \centering
  \includegraphics[width=\columnwidth]{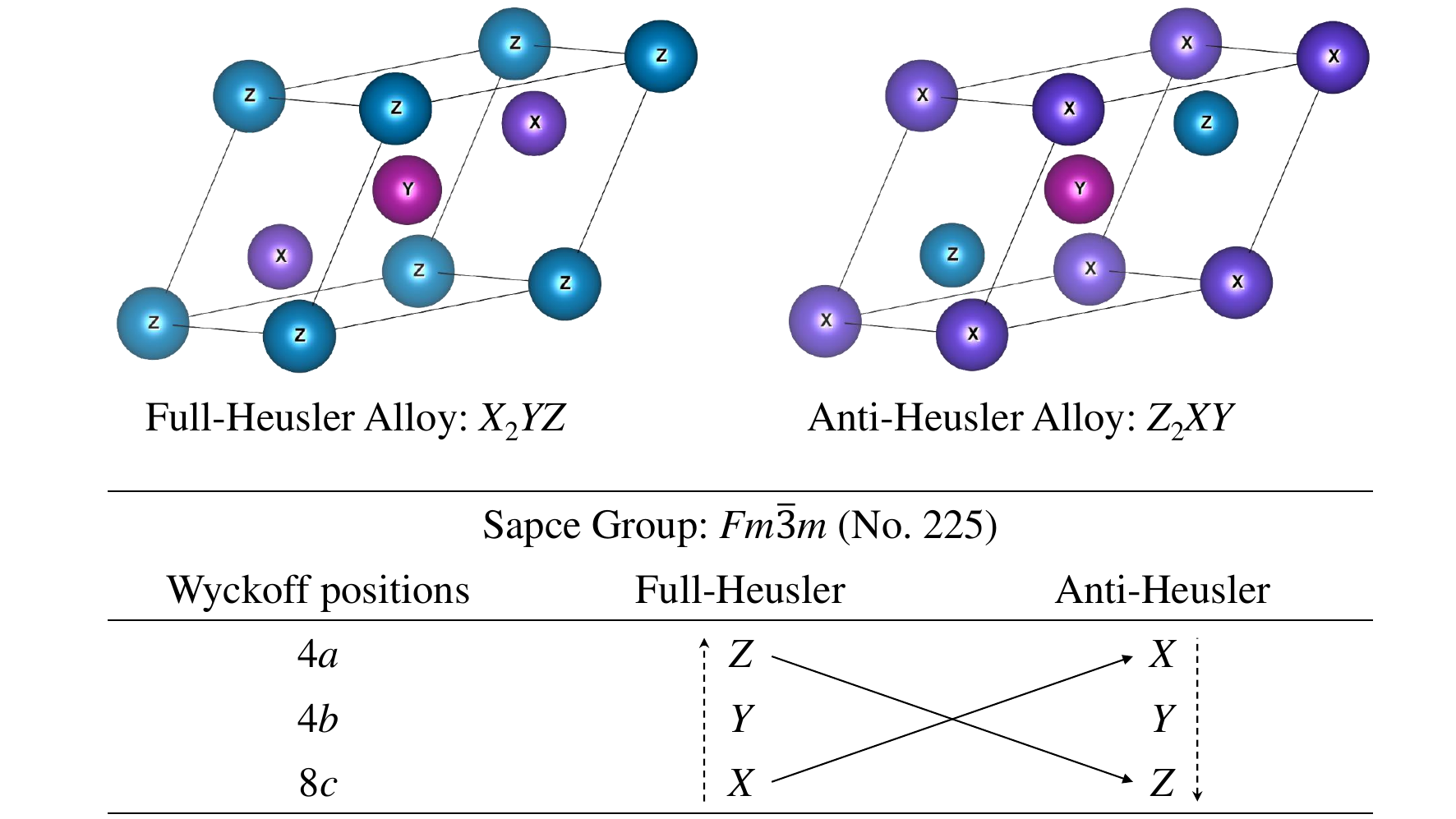}
\label{Fig: anti-Heusler structure}
  \caption{Primitive cell of full-HA and anti-HA. Table describes reversing of atoms ($X$ and $Z$) from full-HA to anti-HA.}\label{Fig:Anti-Heusler structure}
\end{figure}
\begin{figure*}[ht]
  \centering
  \includegraphics[width=0.9\textwidth]{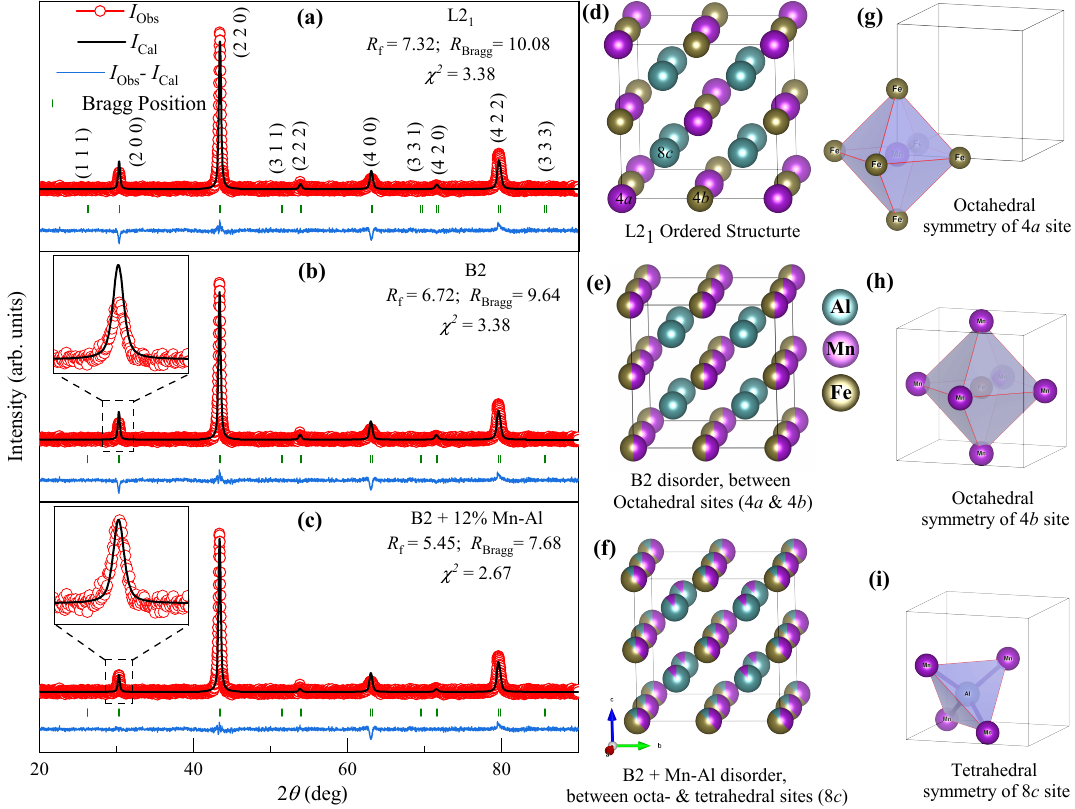}
  \caption{ Rietveld refinement of the X-ray diffraction pattern assuming (a) L2$_{\rm 1}$ ordered structure, (b) B2-type and (c) B2 + 12\% Mn-Al disordered structure, and its associated atomic arrangements are shown in (d), (e), and (f), respectively. 4$a$ and 4$b$ octahedral symmetry sites are represented in (g) and (h), whereas 8$c$ tetrahedral symmetry site is depicted in (i).}
  \label{Fig:XRD}
\end{figure*}
Most of the full-HA having formula unit $X_2YZ$, crystallize in a $fcc$ cubic structure in a space group ${Fm\bar{3}m}$ (No. 225) where $X$, $Y$ and $Z$ atom occupies at 8$c$ (0.25, 0.25, 0.25), 4$b$ (0.5, 0.5, 0.5) and 4$a$ (0, 0, 0), Wyckoff positions respectively, called as L2$_1$-type ordered structure~\cite{GRAF20111}. In contrast, for anti-HA ($Z_2XY$) the atomic arrangements in the ordered structure, also belonging to the same space group, get reversed, $i.e.,$ $Z$ at 8$c$, $Y$ at 4$b$ and  $X$ at 4$a$ Wyckoff sites (Fig.~\ref{Fig:Anti-Heusler structure})~\cite{Samanta2022}. In practice, very few Heusler alloys could be formed in the above-mentioned perfect ordered atomic arrangements. Due to the almost similar electronic structures and atomic sizes of the constituent elements, many of them often belonging to the same row of the periodic table, swap disorder between the atoms in their respective Wyckoff positions is quite a common feature in HA~\cite{PhysRevB.69.144413,10.1063/1.1513216}. The extent of atomic disorder of such compounds is reflected on the intensities of Bragg peaks of diffraction pattern. The scattering factor of Bragg peak with Miller indices ${(hkl)}$ for this structure can be written as~\cite{PhysRevB.99.104429}
\begin{equation}
F_{hkl}= 4(f_{4a}+f_{4b}e^{\frac{\pi}{2}i(h+k+l)}+2f_{8c}e^{\pi i (h+k+l)} )
\label{eq:sf}
\end{equation}
where $f_{4a}$, $f_{4b}$, and $f_{8c}$ are the form factors of the atoms occupying in 4$a$, 4$b$, and 8$c$ sites, respectively. Owing to the $fcc$ lattices, selection rule permits only the Bragg peaks having unmixed Miller indices. Consequently, only three types of distinctive Bragg peaks are allowed and  the scattering factors of these peaks derived from  Eq.~\ref{eq:sf} can be expressed as
\begin{eqnarray}\label{eq:sf1}
  F_{h,k,l \,(all \,odd)} &=& 4 |f_{4a}-f_{4b}| \nonumber\\
  F_{h+k+l=4n+2} &=& 4|(f_{4a}+f_{4b})-2f_{8c}| \\
  F_{h+k+l=4n} &=& 4|f_{4a}+f_{4b}+2f_{8c}|\nonumber
\end{eqnarray}
where $n$ is an integer. Generally, the first three Bragg peaks, appearing in the diffraction spectra, having Miller indices (111), (200) and (220), respectively, satisfies the above three conditions  and generally carry most of the essential information about the nature of atomic disorder in the compound. As Eq.~\ref{eq:sf1} dictates, the (111) peak  intensity depends on the atoms occupying at octahedral sites (4$a$ and 4$b$) only.
In case of a complete intra-symmetry site disorder between atoms at octahedral 4$a$ and 4$b$ sites, known as B2-type disorder, (111) peak intensity would vanish. However, if this disorder occurs between inter-symmetry site, $i.e.,$ in between octahedral  (4$a$/4$b$) and tetrahedral (8$c$) sites, called DO$_3$ disorder, then it affects the relative height of (200) Bragg peak. On the other hand, a totally random disorder between all the Wyckoff sites, known as A2-type disorder, both (111) and (200) peak intensities would reduce in comparison to the intensity of (220) peak, which remains unaffected by this disorder. The extent of chemical disorder in the system can be expressed through two parameters, S and $\alpha$, estimated from the change in ratio of the diffraction peak intensities  $I_{(111)}/I_{(220)}$ and $I_{(200)}/I_{(220)}$  between the experimentally observed data and theoretically generated ordered structure. Quantitatively, one can write $S = ((I_{200}/I_{220})_{Exp.}/(I_{200}/I_{220})_{Th.})^{1/2}$ and $S(1-2\alpha) = ((I_{111}/I_{220})_{Exp.}/(I_{111}/I_{220})_{Th.})^{1/2}$. The well ordered structure yields $S$ = 1 and $\alpha$ = 0 while for B2 and A2-type disorder the order parameter becomes $S$ = 1, $\alpha$ = 0.5 and $S$ = 0, $\alpha$ = 0, respectively\cite{10.1063/1.4959093}.

\subsubsection{X-ray diffraction}

To check the single-phase nature and crystal structure, x-ray diffraction (XRD) measurement on powdered Al$_2$MnFe has been carried out at room temperature. All the XRD peak positions could be indexed considering a cubic unit cell (space group ${Fm\bar{3}m}$ (No. 225)) having lattice parameter, $a=b=c=$ 5.893(1) \AA, indicating single phase nature of the material (Fig.~\ref{Fig:XRD}~(a)). Full Rietveld refinement suggests a well-ordered L2$_1$ -type crystal structure (Fig.~\ref{Fig:XRD}~(d)), where Al atoms are at tetrahedral site 8$c$ (0.25, 0.25, 0.25) (Fig.~\ref{Fig:XRD}~(i)), and the Mn and Fe atoms occupy the octahedral sites, 4$a$ (0, 0, 0), and 4$b$ (0.5, 0.5, 0.5) Wyckoff positions, respectively (Fig.~\ref{Fig:XRD}~(g), (h)). It is noteworthy to mention here that the XRD spectra can be explained well by ordered L2$_1$ structure despite the total absence of any detectable (111) Bragg peak intensity, which is generally considered to be a signature of B2-type atomic disorder. Actually, the fit quality remain invarient even we consider B2-type disorder instead (Fig.~\ref{Fig:XRD}~(b), (e)). To understand this apparent contradiction, one need to revisit Eq.~\ref{eq:sf1}, specifically applicable for Al$_2$MnFe. Placing Mn, Fe, and Al at 4$a$, 4$b$ and 8$c$ sites, respectively, Eq.~\ref{eq:sf1} can be rewritten as
 \begin{eqnarray}
 \label{eq:sfofAl2MnFe}
  F_{(111)} &=& 4|f_{Mn}-f_{Fe}|\nonumber\\
  F_{(200)} &=& 4|f_{Mn}+f_{Fe}-2f_{Al}| \\
  F_{(220)} &=& 4|f_{Mn}+f_{Fe}+2f_{Al}|\nonumber.
\end{eqnarray}

We thus see that (111) peak intensity actually represents the difference between scattering factors of Mn and Fe at 4$a$ and 4$b$ sites, respectively. Since the x-ray scattering factors of Mn (Z=25) and Fe (Z=26) are almost equal, the (111) peak intensity appears to be vanishingly small in the XRD pattern, even for the perfectly ordered L2$_1$-type atomic arrangement. Since B2-type disorder have a discernible effect only on the (111) peak intensity, its absence even in ordered structure pose a great hindrance in identifying the actual atomic arrangement from the XRD pattern alone. Consequently, we have carried out theoretical energy minimization calculation (Sec.~\ref{Theory}), establishing the most stable form is of B2-type disorder.

A closer examination of the XRD fit analysis reveals a noticeable reduction in the experimentally observed intensity of the (200) Bragg peak at 2$\theta \sim30^{0}$  compared to the theoretically generated pattern based on the B2-disorder structural model (inset of Fig.~\ref{Fig:XRD}~(c)). As seen in Eq.~\ref{eq:sfofAl2MnFe}, the scattering factor for (200) Bragg peak is represented by the difference between the sum of total scattering factors at 4$a$ and 4$b$ sites (occupied randomly by Mn and Fe atom) and twice of the scattering factor of atoms at 8$c$ site (Al atom). The experimental observation of lower intense (200) peak suggest an atom swapping between the octahedral 8$c$ site and tetrahedral 4$a$/4$b$ sites. The subsequent Rietveld analysis suggests such inter-symmetry swapping strength to be $\sim$12\%, $i.e$, 1/8 (Fig.~\ref{Fig:XRD}~(c)). We thus find that while a large (50\%) Mn-Fe disorder remains indistinguishable in XRD due to their similar scattering factors, the distinct scattering factor of Al enables the identification of a rather relatively small 12\% disorder. However, the close-by x-ray scattering factors of Fe and Mn again put an hindrance in identifying the swap preferences between the Fe and Mn with Al. Nevertheless, by combining the XRD analysis with insights from theoretical calculations (Sec.~\ref{Theory}) and $^{57}$Fe Mössbauer spectroscopy (Sec.~\ref{Mössbauer spectrometry}), we identify the above-mentioned $\sim$12\% inter-symmetry site swapping as arising from Al-Mn disorder. The resultant atomic arrangement of Al$_2$MnFe are presented in Table~\ref{tab:atom occupancy of Al2MnFe} and the crystal structure is shown in Fig.~\ref{Fig:XRD}~(f), which can be ascribed as B2 with 12\% Mn-Al disorder. Our structural analysis on Al$_2$MnFe is aligned with that of another $Z_2XY$-type anti-Heusler compound, Al$_2$MnCu, where also a similar B2 + 5\% inter-symmetry site disorder was reported ~\cite{PhysRevB.111.174417}.

It may be mentioned here that a similar compound having the same elemental composition has been reported elsewhere in an independent and parallel study, where the crystal structure have been explained considering the B2-type disorder only ~\cite{KHORWAL2025183936, khorwal2024magneticcriticalphenomenalow}. The additional $\sim$12\% Mn-Al disorder that we have estimated from the subtle change in experimental Bragg peak intensity near 2$\theta \sim$ 30$^{\circ}$ vis-$\grave{a}$-vis to that theoretically proposed by B2-type disorder was not considered in the other work. However, the expected mismatch of intensity around the same Bragg peak position is indeed quite discernible in the XRD data presented in Ref.~\cite{khorwal2024magneticcriticalphenomenalow} (evident from the dip in the $I_{obs}-I_{cal}$ at $\sim$30$^{\circ}$), although the signature is not so apparent in the corresponding figure in Ref.~\cite{KHORWAL2025183936} which we believe is due to the limitation of signal-to-noise ratio in the later work ~\cite{KHORWAL2025183936}.

\begin{table}
\centering
\resizebox{\columnwidth}{!}{%
\begin{tblr}{
  cells = {c},
  cells = {c, valign = m},
  cell{2}{1} = {r=2}{},
  vlines,
  hline{1-2,4-5} = {-}{},
  hline{3} = {2-5}{},
}
Site        & {Wyckoff \\site} & (x, y, z)                                 & Element      & Occupancy          \\
Octahedral  & 4$a$               & (0, 0, 0)                                 & {Fe\\Mn\\Al} & {0.5\\0.38\\0.12~} \\
            & 4$b$               & ($\frac{1}{2}, \frac{1}{2}, \frac{1}{2}$) & {Fe\\Mn\\Al} & {0.5\\0.38\\0.12}  \\
Tetrahedral & 8$c$               & ($\frac{1}{4}, \frac{1}{4}, \frac{1}{4}$)      & {Al\\Mn}     & {0.88\\0.12}
\end{tblr}
}
\caption{Atomic occupancy of Al$_2$MnFe, in octa- and tetra-hedral sites.}
\label{tab:atom occupancy of Al2MnFe}
\end{table}

\subsection{Magnetic properties}
\subsubsection{dc magnetization}
\label{dc magnetization}
\begin{figure*}[t]
  \centering
  \includegraphics[width=1\textwidth]{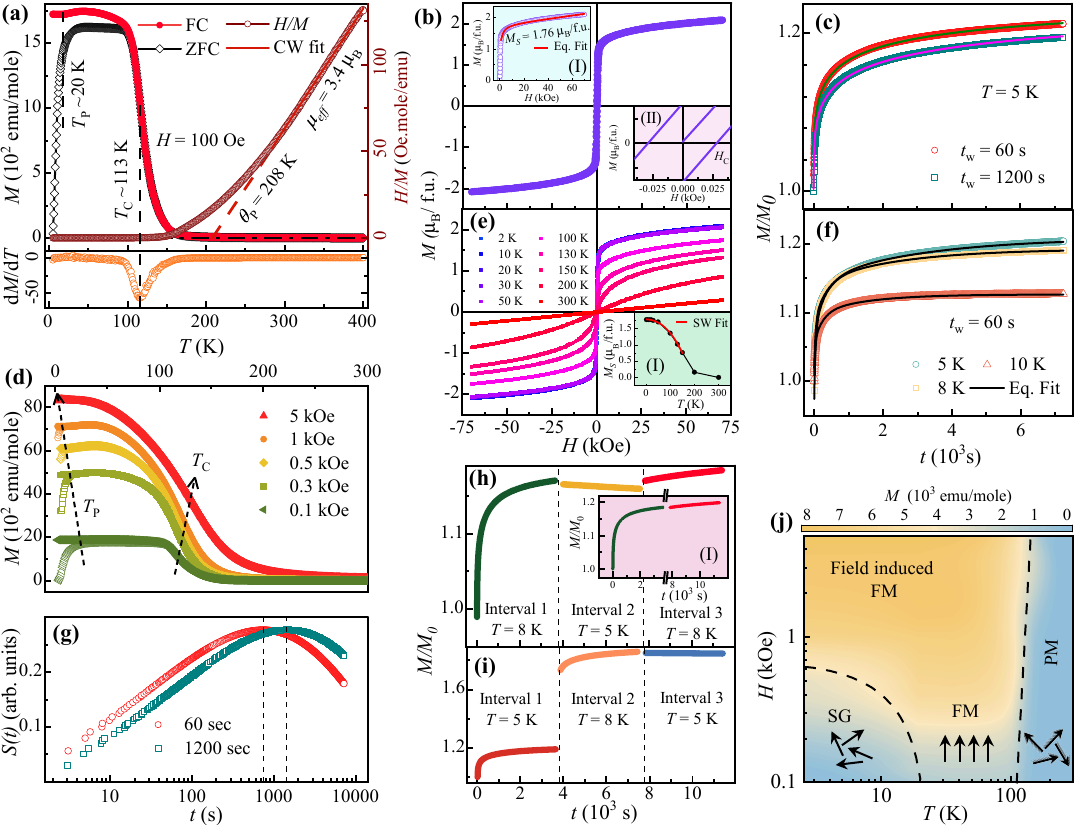}
  \caption{(a) Temperature dependence ZFC and FC magnetization curve under 100 Oe applied field (left), $H/M$ curve (right), d$M$/d$T$ curve (lower panel); (b) Field dependence magnetization curve at 2~K, upper (I) and lower (II) inset shows spontaneous magnetization fitting and coercivity respectively; (c) Wait time dependent magnetic relaxation measured in ZFC protocol at $T$= 5~K (d) $M(T)$ curve under various applied field, where closed symbols are FC curves whereas the open symbols represents ZFC curves. Left and right arrow shows the gradual change of $T_{\rm P}$ and $T_{\rm C}$ respectively. (e) $M(H)$ curve measured at different temperatures, inset represent temperature dependence spontaneous magnetization value, which follows the SW equation. (f) Temperature dependent relaxation with fixed wait time, $t_{\rm w}$= 60~s. (g) Shifting of hump in relaxation rate $S(t)$ due to different wait time. Memory effect measured in (h) intermediate cooling and (i) intermediate heating protocol. Inset (I) of (h) is the merging data of interval 1 and 3 measured in cooling protocol. (j) Magnetic phase diagram of Al$_2$MnFe, obtained from zero-field-cooled (ZFC) magnetization data. The low temperature dashed curve is drawn to show the spin glass (SG) region, where the disordered spins are frozen, depicted by arrows. The ferromagnetic (FM) region is in between the two dashed curve, where spins are arranged in parallel to each other, shown by parallel arrows. The high temperature region above the dashed line is the paramagnetic (PM) region, where the spins are randomly oriented and vibrates due to thermal energy.}
\label{Fig:dc magnetization}
\end{figure*}

To understand the basic magnetic properties of the system, temperature dependent magnetization, $M(T)$, measurement is performed in zero-field-cooled (ZFC) and field-cooled (FC) protocols under various applied magnetic fields in the temperature range of 2-400~K. The $M(T)$ feature under applied magnetic field, $H$ = 100~Oe is shown in Fig.~\ref{Fig:dc magnetization}~(a). The compound exhibits a paramagnetic (PM) to ferromagnetic (FM) phase transition with curie temperature $T_{\rm C}\sim$ 113~K, estimated from the dip of the $dM/dT$ behavior. A bifurcation between the ZFC and FC curves emerges below a thermomagnetic irreversibility temperature $T_{\rm irr}\sim$100~K, accompanied by a magnetization downturn below $T_{\rm P}\sim$20~K. This thermomagnetic irreversibility, along with the reduction in both ZFC and FC magnetization, indicates the destabilization of long-range FM interactions below $T_{\rm P}$. This is also manifested in the magnetic field dependent magnetization behavior, shown in Fig.~\ref{Fig:dc magnetization}~(d). Upon increasing applied field strength, the curve gets broadened around $T_{\rm C}$ yielding upwards shift in temperature for the dip in $dM/dT$, typically observed for a long-range PM-FM transition. However, the shift of the hump at $T_{\rm P}$ towards lower temperatures with increasing applied field suggest the existence of metastable magnetic spins at low temperature region in the compound.

The inverse susceptibility data exhibits a linear temperature dependence only above 280~K, indicating the existence of short range magnetic interaction extending far above $T_{\rm C}$ $\sim$113~K. Above 280~K, the magnetic susceptibility can be described by the Curie-Weiss (CW) behavior, $\chi = C/(T-\theta_{\rm CW}$), where $C = N_ A\mu_{eff}^2\mu_B^2/3K_B$, is the Curie constant and $\theta_{\rm CW}$ is the paramagnetic Curie-Weiss temperature. Fig.~\ref{Fig:dc magnetization}~(a)~(right scale) presents the inverse of magnetic susceptibility and the corresponding CW fit for $T$ $>$ 280 K, yielding  $\theta_{\rm CW}\sim$208~K and $\mu_{\rm eff}\sim3.4~\mu_{\rm B}$. The high and positive $\theta_{\rm CW}$ value suggests a dominating ferromagnetic interaction in the compound.

The magnetic field dependence of isothermal magnetization $M(H)$ measured at $T = 2$~K is shown in Fig.~\ref{Fig:dc magnetization}~(b), while Fig.~\ref{Fig:dc magnetization}~(e) illustrates the behavior across all measured temperatures. At 2~K, $M(H)$ increases sharply in the low field region with a low coercivity of $H_{\rm C}\sim$ 20~Oe (Fig.~\ref{Fig:dc magnetization}~(b), inset (II)), a characteristic typically observed in soft ferromagnetic materials. Additionally, the non-saturating trend at higher applied fields results in an overall ‘S’-shaped curve, suggesting that not all spins are aligned ferromagnetically. Such behavior is often associated with competing magnetic interactions, where a magnetically glassy phase coexists with ferromagnetic order, leading to a similar magnetization response ~\cite{ PhysRevB.108.054405}. Accordingly, the $M(H)$ data has been analyzed using an approach to saturation law (Fig.~\ref{Fig:dc magnetization}~(b), inset (I)), $M= M_S \left(1-\frac{A}{H}- \frac{B}{H^2}\right)+\chi_h H$, where $M_S$ represents the spontaneous magnetization, the parameters $A$ and $B$ account for the effects of structural defects and magneto-crystalline anisotropy, respectively, and the term $\chi_h$ represents the high-field susceptibility~\cite{principles}. The spontaneous magnetic moment of the compound is thus obtained as $M_{\rm S} \sim$1.76$~\mu_{\rm B}$/f.u., indicating that ferromagnetic order remains dominant, despite the presence of signatures associated with a magnetically metastable phase at low temperatures. This feature also suggests that the magnetic ground state of the compound is magnetically inhomogeneous, where a dominating long-range ordered FM state coexists with a short-range ordered glassy phase. The temperature-dependent values of $M_{\rm S}$, obtained by fitting with the approach to saturation law to different $M(H)$ curves detailed above, are shown in the inset of Fig.~\ref{Fig:dc magnetization}~(e) that follow the magnetization trend which can be described by the spin-wave equation, $M_S(T)=M_S(0)(1-AT^{3/2}-BT^{5/2})$, where $M_{\rm S}(0)$, the spontaneous magnetization value at $T=0$~K, is determined as 1.768(7) $\mu_{\rm B}$/f.u., A and B are estimated as 7.67$\times10^{-5}$ and 1.59$\times10^{-7}$ respectively, which are typically observed in ferromagnetic systems~\cite{PhysRevB.50.9308}. The overall nature of the $M(H)$ curves remains largely unchanged up to 50~K. As the temperature is increased further to the Curie temperature, $T_{\rm C}$ $\sim$113~K, the saturation magnetic moment starts to decrease gradually. Noteworthy, the presence of short-range magnetic interaction, as interpreted from the temperature-dependent non-linear behaviour of inverse susceptibility in the temperature range, $T_{\rm C}<T<$280~K (Fig.~\ref{Fig:dc magnetization}~(a)), also influences the isothermal magnetisation behaviour to be magnetic-field dependent non-linear in the same temperature range (Fig.~\ref{Fig:dc magnetization}~(e)). Such short-range interaction may have many different origins, including the formation of superparamagnetic clusters~\cite{pssa.201127016}, the magnetic precursor effect~\cite{PhysRevB.76.014401}, etc. In a typical superparamagnetic system, the $M(H)$ isotherms are expected to collapse into a single universal curve when the reduced magnetization ($M/M_S$) is plotted as a function of $H/T$~\cite{PhysRevB.98.205130, PhysRevB.78.064401}. However, the resultant plot fails to conform superparamagnetic state, even if we consider the magnetisation to have a partial paramagnetic contribution in addition to the short-range magnetic interaction. The ac-susceptibility results presented later (Sec.~\ref{ac susceptibility}), also rule out superparamagnetic contribution to magnetism in this temperature range.

\subsubsection{Magnetic relaxation and memory effect}
\label{Magnetic relaxation and memory effect}

The decrease in $M(T)$ below $T_{\rm P}$ in both ZFC and FC modes indicates a destabilization of the magnetically ordered spin structure. The presence of metastable spins is further examined by studying the non-equilibrium dynamics of the system through the wait-time- and temperature-dependent magnetic relaxation as well as magnetic memory effect measurements~\cite{principles,RevModPhys.58.801}. In the wait-time dependent magnetic relaxation measurement, the sample is cooled to a desired temperature, $T(<T_{\rm P})$, from the true paramagnetic region ($>$300~K for Al$_2$MnFe) in the absence of any external field. After waiting for a specific time, $t_{\rm w}$ at $T$, a small field is applied and the evolution of magnetization $M(t)$ is recorded over an extended period of time, $t$. Fig.~\ref{Fig:dc magnetization}~(c) illustrates the $t_{\rm w}$-dependent relaxation behavior of Al$_2$MnFe at $T=$ 5~K for $t_{\rm w}$= 60 and 1200~s.

In the case of a magnetically glassy material, the spins are in a non-equilibrium state when the system is cooled across the blocking temperature in the absence of any external field. The non-equilibrium state of the magnetic spins cast an imprint on the magnetic relaxation data where the inflection point strongly depends on the wait time, $t_{\rm w}$~\cite{PhysRevB.94.104414}. The inflection point is manifested as a peak in the magnetic viscosity, $S(t)=\frac{1}{H}.\frac{dM(t)}{d(log(t))}$. Accordingly, $S(t)$ is plotted as a function of time for $T=$ 5~K (Fig.~\ref{Fig:dc magnetization}~(g)). The curve shows a prominence peak, which shifts towards a higher time for a longer wait time. This is known as the aging phenomenon, which is a signature of glassy phase and is attributed to the non-equilibrium dynamics of domain growth~\cite{PhysRevB.90.024421,PhysRevB.108.054430}.

Subsequently, we have further extended the study of relaxation behavior for several temperatures below $T_{\rm P}$, but with fixed wait time, $t_{\rm w}$= 60~s (Fig.~\ref{Fig:dc magnetization}~(f)). The relaxation curve is best described by exponential function with three relaxation time constants,
\begin{equation}{\resizebox{\hsize}{!}{$M(t)= M_0-M_1exp[-(t/\tau_1)]-M_2exp[-(t/\tau_2)]-M_3exp[-(t/\tau_3)]$}}
\label{eq.Relaxation}
\end{equation}
\noindent where $M_0$ is the time-independent intrinsic magnetization and $M_1$, $M_2$, and $M_3$ are time-dependent fractions of magnetization which relax with time-constants, $\tau_1$, $\tau_2$, and $\tau_3$~\cite{PhysRevB.74.184430,
Dasgupta2017}. The presence of three different values of $\tau$ indicate that the system has at least three distinct clusters with multiple energy barriers. The obtained relaxation time constants $\tau_1$, $\tau_2$, and $\tau_3$ at 5~K are 43, 401, and 3086~s; at 8~K are 37, 299, and 2377~s; at 10~K these constants become 26, 244, and 1448~s. These reducing time-constants with increasing temperature indicate the gradual decrease of stiffness of spin relaxation due to thermal energy and is a common feature for spin/cluster glass system~\cite{SAHA2021159465}.

The magnetic memory effect is another key non-equilibrium dynamical feature of glassy systems~\cite{PhysRevLett.91.167206,principles}. To investigate this property in Al$_2$MnFe, the sample was first cooled to 8~K ($<T_P$) from the paramagnetic temperature ($\sim$ 300~K) under a small applied field of 100~Oe. The magnetization $M(t)$ was then recorded for $t_1$= 3600~s (interval 1). Following this, the sample was rapidly cooled to 5~K without altering the applied field, and $M(t)$ was measured again for $t_2$= 3600~s (interval 2). Finally, the sample was reheated back to 8 K, and $M(t)$ was recorded for another $t_3$= 3600~s (interval 3).

The resulting $M(t)$ curve, shown in Fig.~\ref{Fig:dc magnetization}~(h). The inset~(I) of Fig.~\ref{Fig:dc magnetization}~(h), reveals that the curves corresponding to intervals 1 and 3 are nearly continuous, indicating that the system effectively ``remembers" its prior state during the cooling period and resumes from its initial state upon reheating. This strong correlation suggests a robust memory effect in the system. It is worth noting that such a seamless matching of data is not observed when the inverse temperature protocol is applied and the system rejuvenates, as shown in Fig.~\ref{Fig:dc magnetization}~(i).

The two seemingly different results can be explained by two well-established theories: (i) the droplet model~\cite{PhysRevB.38.373, WLMcMillan1984} and (ii) the hierarchical model~\cite{VSDotsenko1985}. According to the droplet model, only a single spin configuration exists at each temperature. In contrast, the hierarchical model proposes a multi-valley free energy landscape at any given temperature. If the system follows the hierarchical model, it should exhibit a memory effect only in the cooling protocol, whereas the droplet model predicts memory effects in both cooling and heating protocols. Our observations of the different memory effects suggest the presence of a multi-valley free energy landscape, aligning well with the temperature-dependent magnetic relaxation results presented earlier~\cite{PhysRevB.107.184408,PhysRevB.99.174410,PhysRevB.94.104414}.

\subsubsection{ac susceptibility}
\label{ac susceptibility}
\begin{figure*}[ht]
  \centering
  \includegraphics[width=1\textwidth]{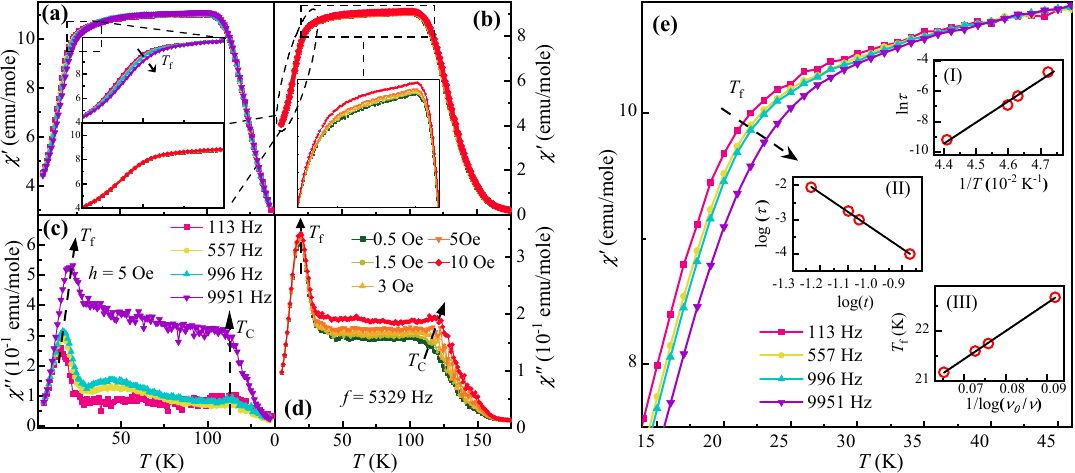}
  \caption{Frequency dependent (a) real ($\chi^{\prime}$) and (c) imaginary $\chi^{\prime\prime}$ part of ac-susceptibility measured at fixed excitation field $h$= 5 Oe.The real part curve get bifurcated below $T_{\rm f}$, which also increase to higher temperature (in imaginary part also) due to the applied higher frequency, shown in inset. (b) and (d) depicts excitation field dependence $\chi^{\prime}$ and $\chi^{\prime\prime}$ respectively, measured at fixed frequency $f$= 5329 Hz. Here the curve bifurcated below $T_{\rm C}$ and subsequently merges after $T_{\rm f}$. The arrow shows $T_{\rm f}$ does not changes due to the variation of excitation field. (e) Blown up of the frequency dependent real part of ac susceptibility curve. Inset (I) Arhenius law fitting, (II) critical scaling theory fitting and (III) Vogel Fultcher theory fitting of the freezing temperatures with respect to the applied frequencies.}
\label{Fig:AC}
\end{figure*}

The Non-equilibrium dynamics of spins, presented above, emphatically establishes the formation of disordered spin structure likely to have driven by a short range exchange interactions. Since depending on the nature of such short range interactions, the disordered spin-structure can be classified in different sub-classes, $e.g.,$ spin-glass, cluster-glass, super-paramagnetic, $etc.,$ frequency dependent ac magnetic susceptibility could be of great help in distinguishing these features~\cite{PhysRevB.6.4220}. Further, the variation of excitation strength of the ac magnetic field can also be utilised to establish the long range magnetic order~\cite{balanda2013ac,Topping2019}. Since Al$_2$MnFe exhibits signature of ferromagnetic order at $\sim$113~K and metastable magnetic spins below $\sim$20 K, we have carried out ac susceptibility measurements at different frequencies as well as in the presence of different driving field values as a function of temperatures (Fig.~\ref{Fig:AC}). The time ($t$) varying field $H(t)$ can be expressed as: $H(t)= H_0 + h cos(2\pi ft)$, where $H_0$ is a static dc field, $h$ is amplitude of the driving field and $f$ is the frequency. To probe the magnetic phases of Al$_2$MnFe, the measurements have been performed in two protocols where dc bias field ($H_0$) is set at zero throughout the experiment. First, frequency ($f$) dependent ac susceptibility data with a constant $h$ (= 5~Oe) in the temperature range 5-150~K, has been recorded to substantiate the glassy nature of the system. In second experiment, the long-range ferromagnetic phase has been probed by varying the driving amplitude, $h$, with fixed $f$ (= 5329~Hz).

The frequency and driving amplitude dependent real part of ac susceptibility curve, $\chi'$, of the studied anti-Heusler, Al$_2$MnFe is shown in Fig.~\ref{Fig:AC}~(a) and (b), respectively. For both the measurements, $\chi'$ show a table-like plateau, confined between Curie temperature, $T_{\rm C}\sim$ 113 K, in the high-temperature side and $T_{\rm P}\sim$20 K in the low-temperature side, which is consistence with dc magnetization result presented earlier (Sec.~\ref{dc magnetization}). On close scrutiny, however, discernible differences are quite evident between the frequency dependent, $\chi'(T, f)$, and driving field dependent, $\chi'(T, h)$, data. For $T < T_{\rm P}, ~\chi'(T, f)$ traverse differnt paths whereas $\chi'(T, h)$ remain indistinguishable. On the other hand, for $T_{\rm P}< T <T_{\rm C}$, the feature is opposite for  $\chi'(T,f)$ and  $\chi'(T,h)$ (Fig.~\ref{Fig:AC}). In other words, $\chi'$ curves in the plateau region ($T\sim$90-30~K) does not response in the change of $f$ whereas, those diverges due to the variation of $h$. Contrastingly, $\chi'$ curves split in the application of different frequencies in the low temperature region ($<T_{\rm P}$) whereas, it remains unchanged upon changing of driving field, $h$. These contrasting behavior of $\chi'$ in two measurement protocols detect existence of two magnetic phases in the system. The field dependence phenomena in the temperature region $T_{\rm P} < T < T_{\rm C}$ indicates the ferromagnetic phase whereas in case of $\chi'(T,f)$, the temperature below which it start to fall ($T_{\rm P}$) shifts towards higher temperature with increasing frequency substantiating the magnetically glassy feature in Al$_2$MnFe. Hence onwards, this phase transition temperature, $T_{\rm P}$, can thus be assigned as the freezing temperature, $T_{\rm f}$, of the frozen spin state.

The imaginary part of ac susceptibility data, $\chi''$, measured in both the protocols are depicted in Fig.~\ref{Fig:AC}~(c) and (d). The non-zero value of $\chi''$, measured in both protocols, obtained below $T_{\rm C}$, suggesting formation of ferromagnetic domains~\cite{Topping2019}. Ferromagnetism is further confirmed in the $\chi''(T, h)$ measurement, where the peak position ($\sim T_{\rm C}$) shifts with increasing value of driving field, but remain unchanged with the variation of frequency in $\chi''(T, f)$, substantiating the absence of superparamagnetic clusters, corroborating with the analysis of $M(H)$ data. On the other hand, the peak around $T_{\rm f}$ display opposite character, $i.e.$, the peak position remains invariant in $\chi''(T, h)$, but shifts in $\chi''(T, f)$ measurement which is the characteristic of magnetically metastable phase~\cite{PhysRevB.27.3100}. Together, both the features substantiate the magnetically re-entrant glassy nature of Al$_2$MnFe.

For a better classification of the metastable character at low temperature($<T_{\rm f}$), $i.e.$, whether the system become reentrant spin-glass, cluster-glass or superparamagnetic, one needs to assess the extent of the temperature shift with frequency. The relative shift of temperature with respect to frequency shift per decade is expressed by Mydosh parameter, which is defined as~\cite{principles}
\begin{equation}\label{eq:Mydosh parameter}
  K=\frac{\triangle T_f}{T_f\,log_{10}(\triangle f)}
\end{equation}
where $f$ is frequency and $T_{\rm f}$ is spin freezing temperature. Mydosh parameter, $K\sim$0.001 found for canonical spin glasses~\cite{PhysRevB.23.1384}, $\sim$0.01 is reported for spin-cluster-glass compounds~\cite{PhysRevB.106.224427,PhysRevB.107.184408} and in the range of 0.1-0.28 signifies superparamagnet~\cite{principles}. For Al$_2$MnFe, $K$ has been estimated as $\sim$0.007 which lies on the border of the cluster-glass and canonical spin-glass regimes.

Many different empirical laws, $viz.,$ (i) Arrhenius law, (ii) critical scaling approach, and (iii) Vogel-Fulcher law that utilize the information of shifting of $T_{\rm f}$ with frequency are often used to identify the metastable phase. We start with Arrhenius law, which is generally applicable for superparamagnetic system and defined as
\begin{equation}\label{eq.Arrhenius law}
  \tau = \tau_0\,exp\,(E_a/K_BT)
\end{equation}
where $\tau$ denotes the inverse of frequency, $\tau_0$ is single spin relaxation time, and $E_a$ is the activation energy~\cite{balanda2013ac,Topping2019}. For a typical superparamgnetic system, $\tau_0$ is found in the range of 10$^{-9}$ $-$ $10^{-12}$~s~\cite{principles}. The estimated value of $\tau_0$ for Al$_2$MnFe, is $\sim10^{-31}$~s (Fig~\ref{Fig:AC}~(e)~inset I). The obtained unphysically small value of single spin relaxation time, $\tau_0$, negates the possibility of superparamgnetic phase in the system.

As the Arrhenius law fails to describe the frequency dependence of $T_f$, critical scaling theory has been applied. As per this law:
\begin{equation}\label{eq.critical scaling formula}
  \tau = \tau_0 \left(\frac{T_f-T_{SG}}{T_{SG}}\right)^{-z\nu}
\end{equation}
where $T_{SG}$ is the spin glass temperature at zero frequency, $z\nu$ is critical exponent of correlation length, $\xi= (T_f/T_{SG}-1)^\nu$~\cite{RevModPhys.49.435}. For a spin glass state, $z\nu$ value is found in the range of 4 to 12~\cite{PhysRevB.108.054405}. The $\tau_0$ value is reported for  canonical spin glass systems within 10$^{-13}$ $-$ 10$^{-12}$ s, and for spin-cluster-glasses the value lies in the range of 10$^{-11}$$-$10$^{-4}$ s~\cite{PhysRevB.94.104414}. Beside that, $\tau_0$ becomes higher for superparamagnetic systems~\cite{PhysRevB.86.064412}. The best fit is achieved for Al$_2$MnFe with $z\nu\sim$ 5.42 and $\tau_0\sim$10$^{-9}$~s (Fig.~\ref{Fig:AC}~(e)~inset (II)). The value of $z\nu$ is in the range that typically found for spin-glasses. However, the $\tau_0$ is higher than that expected in the canonical spin glass regime but lies in the range of spin-cluster-glass systems.

Another approach is to utilize the Vogel-Fulcher law~\cite{tholence1980frequency}:
\begin{equation}\label{eq.vogel Fulcher}
  \tau = \tau_0\,exp\,\left[\frac{E_a}{K_B(T_f-T_0)}\right]
\end{equation}
where, $T_0$ is the Vogel-Fultcher temperature, representing the interaction between the spins.  The value of  $E_a/K_BT_0$ defines the strength of the exchange interaction, which becomes close to unity for canonical spin-glass systems and generally found to be larger that unity for a spin-cluster-glass~\cite{principles}. The obtained parameters for the studied compound  are $E_a/K_B$= 56.39~K, $T_0$= 17.49~K, and $E_a/K_BT_0$= 3.22 (Fig.~\ref{Fig:AC}~(e)~inset (III)). The obtained non-zero value of $T_0$, indicates the formation of clusters. Further, this value is close to the spin freezing temperature, suggesting Ruderman-Kittel-Kasuya-Yosida (RKKY)interaction is dominating in this compound~\cite{PhysRevB.85.014418}. The estimated ratio, ($E_a/K_BT_0 \sim$3.22), is indicating towards the formation of spin-cluster-glass in the system~\cite{PhysRevB.94.104414}.

Combining all the above mentioned applicability of various laws involving frequency dependent spin freezing temperature shift, we find that the possibility of superparamagnetic state is negated by non-applicability of Arrhenius law in this system. While the Mydosh parameter and critical scaling parameter can not distinguish between canonical and spin-cluster glass phases, Vogel-Fulcher law rather categorically substantiates that the studied compound, Al$_2$MnFe, is a cluster-glass system. High temperature ferromagnetic order is also reflected in the field-dependent behaviour of the ac susceptibility peak close to 113~K, firmly establishing the system as reentrant spin-cluster-glass system.

Consequently, the magnetic phase diagram  has been constructed for Al$_2$MnFe, utilizing dc and ac magnetization data (Fig.~\ref{Fig:dc magnetization}~(j)). The phase diagram depicts that three magnetic phase exist in the compound depending on the applied field, $H$ and temperature, $T$. The magnetically frozen spin-cluster-glass state appears below $T_f\sim$ 20~K and $H\sim$ 600~Oe, upon applying higher field the compound become field induced ferromagnet. The intrinsic ferromagnetic phase exist in the temperature region 20~K$<$T$<$113~K, whereas the compound become paramagnet at higher temperature.

\subsection{Mössbauer spectrometry}
\label{Mössbauer spectrometry}
\begin{figure*}[t]
  \centering
  \includegraphics[width=1\textwidth]{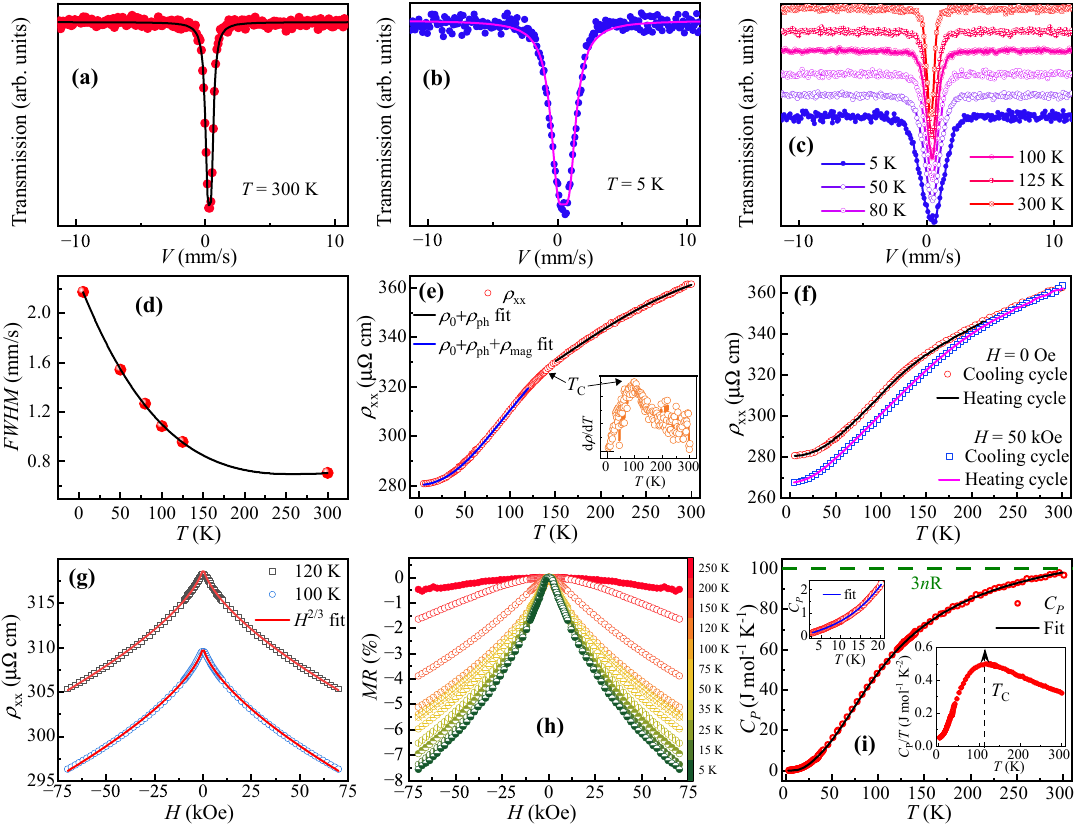}
  \caption{Mössbauer spectrum measured at (a) $T$=300~K, (b) $T$ = 5~K, (c) at six different temperatures in between 5-300~K,(d) The points shows the temperature evolution of full width half maxima (FWHM) and the line is the guide to eye.(e) Temperature dependent zero field longitudinal resistivity ($\rho_{xx}(T)$, measured in heating cycle. Inset shows the change of slope at $T_{\rm C}$. (f) $\rho_{xx}(T)$ at $H=$ 0 and 50~kOe measured in both heating and cooling cycle. (g) Field dependent longitudinal resistivity ($\rho_{xx}(H)$) at $T$= 100 and 120~K. (h) Magnetoresistance (\%) data, $\rho_{xx}(H)$ of various temperatures in the range of 5-250~K . (i) Heat capacity curve of Al$_2$MnFe. Inset show the low temperature fitting with $\gamma T+ \beta T^3$.}\label{Fig:Mossbauer-resistivity}
\end{figure*}

The different dc- and ac-magnetisation measurements presented above establishes the presence of moment carrying atoms beyond any doubt. Since Al$_2$MnFe contains two potentially moment carrying atoms, $viz.,$ Mn and Fe, for a deeper understanding of magnetic properties it is necessary to estimate their relative contribution of magnetisation in this material. Since $^{57}$Fe Mössbauer spectroscopic measurements are known to provide the role of Fe atoms, we have carried out the same measurements at different temperatures in the range of 5-300~K. The line width of Mössbauer spectrum at 300~K found to be a little broader than that expected for paramagnetic Fe atoms in the isotropic environment but can be explained well by considering quadrupolar splitting (Fig.~\ref{Fig:Mossbauer-resistivity}~(a)). This splitting appears when the nucleus is subjected to an electric field gradient, which is found in a non-symmetric/ non-cubic crystal environments, where due to the interaction between the electric field gradient and the nuclear quadrupole moment splits the excited $I$ = 3/2 state into a doublet~\cite{blundell2001magnetism}. Although the XRD measurement on Al$_2$MnFe presented earlier can not distinguish between L2$_1$ and B2-structures, the presence of quadrupolar splitting in $^{57}$Fe Mössbauer spectrum rather suggests non-ordered/ non-symmetric environments and hence indicating towards B2-type disorder. As the temperature is gradually lowered below Curie temperature, the spectra get broader (Fig.~\ref{Fig:Mossbauer-resistivity}~(c) and (d)), but the well-defined sextets expected for moment carrying Fe atoms~\cite{PhysRevB.108.245151} are missing in all the spectra even at 5~K (Fig.~\ref{Fig:Mossbauer-resistivity}~(b)). The absence of sextet spectra rules out the contribution of Fe spins in magnetic ordering, thereby unambiguously establishing that the magnetic moment arises solely from the Mn atoms. The gradual broadening of spectra at lower temperature can be explained by considering the transferred hyperfine field generated by ordering of Mn atoms~\cite{PhysRevB.108.045137}. Our theoretical analysis on the structural as well as magnetic aspects of Al$_2$MnFe presented later (Sec.~\ref{Theory}) found to corroborate the Mössbauer spectroscopic result.

\subsection{Transport properties}
\subsubsection{Electrical resistivity and magnetoresistance}
\label{Electrical resistivity and magnetoresistance}

Many members of HA family are also known to exhibit a multitude of different electrical transport properties, including half-metalicity, spin-gapless semiconducting feature, topological insulating character, $etc.$ Often, the $d$-electrons which are close to the Fermi level, play crucial roles in determining the respective material characteristics. Since the anti-HA has much less $d$-electrons in comparison to other (full-, inverse-, half-, or quaternary-) Heuslers, it would be interesting to investigate the transport properties of Al$_2$MnFe by different methods, $e.g.$, resistivity and magnetoresistance, Hall measurements, heat capacity, $etc.,$ which may also make an imprint of magnetic characteristics of the compound.

The electrical resistivity, $\rho(T)$, of Al$_2$MnFe in the temperature range 5-300~K has been presented in Fig.~\ref{Fig:Mossbauer-resistivity}~(e). The temperature coefficient of resistivity (d$\rho$/d$T$) remains positive throughout the temperature range, suggesting metallic character with the residual resistivity ratio (RRR)$\sim$1.29. The magnetic ordering is also reflected in this transport behaviour as d$\rho$/d$T$ exhibit a distinct peak close to Curie temperature (inset of Fig.~\ref{Fig:Mossbauer-resistivity}~(e)), indicating non-equivalent scattering process in the ferromagnetic ($T<T_{\rm C}$) and paramagnetic region. The $\rho(T)$ behaviour in the paramagnetic region are primarily governed by phonon contribution and generally expressed as, $\rho(T)=\rho_0+\rho_{ph}(T)$, where $\rho_0$ represents the scattering of conduction electron with lattice defects, which is temperature-independent whereas $\rho_{ph}$ is the phonon contribution, appears due to the lattice thermal vibration and can be expressed as $\rho_{ph}(T)=A\left(\frac{T}{\theta_D}\right)^5\int_{0}^{\frac{\theta_D}{T}} \frac{x^5}{(e^x-1)(1-e^{-x})} \,dx$~\cite{rossiter1991electrical}. On the other hand, below the magnetic ordering, magnon contribution is included in addition to the phonon contribution, $\rho(T)=\rho_0+\rho_{ph}(T)+\rho_{mag}(T)$, where the magnon term is generally considered to be quadratic in temperature ($\rho_{mag}(T)\sim BT^2$)~\cite{PhysRevLett.110.066601}.

The resistivity measurement has been further extended by carrying out the measurement in both heating and cooling cycle as well as measuring under a large magnetic field of 50~kOe (Fig.~\ref{Fig:Mossbauer-resistivity}~(f)). The absence of any irreversibility suggest the absence of any discernible structural transformation~\cite{PhysRevB.77.224440}. On the other hand, $\rho(T)$ exhibits a considerable lowering of value below $\sim$260~K when an external magnetic field of 50~kOe is applied. The reduction in $\rho(T)$) under the application of magnetic field suggests a likely presence of magnon much above its Curie temperature ($\sim$113~K). In this context, if we would like to correlate $\rho(T, 50~kOe)$ behaviour with magnetic susceptibility, we may find that the inverse susceptibility start to deviate from linear character around the same temperature (Fig.~\ref{Fig:dc magnetization}~(a)) from where $\rho(T, 50~kOe)$ also starts to deviate from its corresponding zero-field measurement (Fig.~\ref{Fig:Mossbauer-resistivity}~(f)). The deviation from linearity in inverse susceptibility is a reflection of presence of short range correlation extending up to $\sim$260 K. When the magnetic field is applied, this short range correlation is responsible for the reduced magnon scattering, resulting lower resistivity values.

In order to better understanding the field dependency, the resistivity measurement is also carried out with the variation of magnetic field in the range of -70 to +70~kOe at different temperatures in between 5-250~K. In the vicinity to $T_{\rm C}$, at 100 and 120~K, the resistivity curve follows $H^{2/3}$ dependence (Fig.~\ref{Fig:Mossbauer-resistivity}~(g)) suggesting the resistivity can be attributed to the $s-d$ scattering of the conduction electrons, which is a typical feature of a ferromagnetic material~\cite{MRJPSJ,PhysRevB.109.134428,PhysRevB.66.024433}. Additionally, as discussed in previous paragraph, negative magnetoresistance (MR) is also confirmed by isothermal resistivity curves. It indicates that the spin fluctuation is reduced by the applied field, resulting lower resistivity value. The MR value is generally quantified with, $MR(H)= \frac{\rho(H)-\rho(0)}{\rho(0)}\times100$. The MR(\%) values at different temperatures in 5-250~K regime in the field range of -70 to +70~kOe is shown in Fig.~\ref{Fig:Mossbauer-resistivity}~(h).
The highest estimated value of MR at 5~K in application of 70~kOe is found to be -7.5\% which decreases with increasing temperature suggest the effect of applied field is reduces in higher temperatures as expected in a (ferro-) magnetic material~\cite{PhysRevB.55.R8650}.

\subsubsection{Anomalous Hall measurement}
\label{Anomalous Hall measurement}
\begin{figure*}[ht]
  \centering
  \includegraphics[width=1\textwidth]{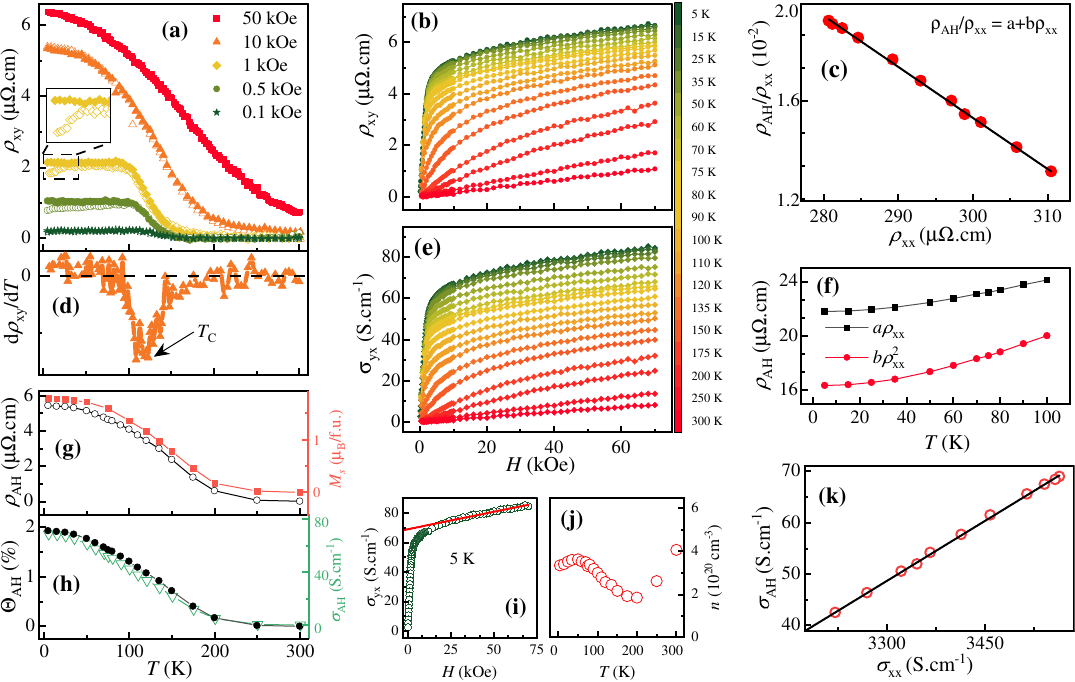}
  \caption{(a) Temperature dependent Hall resistivity ($\rho_{xy}(H)$) at various fields in between 0.1-50~kOe. The open symbols are ZFC and the closed symbols are FC data. (b) Field dependent Hall resistivity ($\rho_{xy}(H)$) at various temperatures in between 5-300~K. (c) Linear fit of  $\rho_{AH}(T)/\rho_{xx}(T)$ vs $\rho_{xx}(T)$ to obtain intrinsic and extrinsic contribution. (d)$d\rho_{xy}/dT$ curve. Arrow pointing out the $T_{\rm C}$. (e) Transverse conductivity extracted from transverse and longitudinal resistivity. (f) Skew scattering ($a\rho_{xx}$) and the combination of side jump and intrinsic ($b\rho_{xx}^2$) contribution to anomalous Hall value. (g) Temperature dependence anomalous Hall, $\rho_{AH}$ and spontaneous magnetization, $M_S$ and (h) anomalous Hall angle, $\Theta_{\rm AH}$  and  anomalous Hall conductivity, $\sigma_{\rm AH}$. (i) Linear fitting of $\rho_{xy}(H)$ to obtain anomalous $\rho_{AH}$ value at 5~K, (j) carrier concentration at different $T$. (k) Linear dependence of anomalous Hall conductivity with respect to longitudinal conductivity.}
  \label{Fig:Hall}
  \end{figure*}
In a ferromagnetic sample normally two additive Hall contribution is found. One is ordinary Hall effect because of the Lorentz force exerted on the conduction electron  due to the application of magnetic field perpendicular to the electric field. Another is anomalous Hall effect, originated due to the spontaneous magnetization~\cite{RevModPhys.82.1539}. To substantiate ferromagnetism in Al$_2$MnFe, the temperature and field dependent transverse Hall resistivity ($\rho_{xy}$) is measured in -70 to +70~kOe at temperature range 5-300~K (Fig.~\ref{Fig:Hall}). The temperature dependent Hall resistivity data, $\rho_{xy}(T)$ is shown in Fig.~\ref{Fig:Hall}~(a), which is similar to the magnetization, $M(T)$, curves (Fig.~\ref{Fig:dc magnetization}~(b)), supports the ferromagnetic behaviour of the compound. Further, the dip of the $d\rho_{xy}/dT$ curve at $\sim$113~K, corroborated well with the obtained $T_{\rm C}$ from the magnetization measurements. Additionally, the thermal irreversibility in ZFC and FC $\rho_{xy}(T)$ data supports the low temperature glassy nature (inset of Fig.~\ref{Fig:Hall}~(a)). These feature is observed in many other spin glass systems as well~\cite{PhysRevLett.93.246605,Pureur2004}. Hence, the evolution of magnetic phases of Al$_2$MnFe, from para- to ferro-mgnetic and finally into spin-glass state at lower temperature, are elucidated in transverse Hall resistivity.

The field dependence Hall resistivity is further measured to parse the different contribution of normal and anomalous Hall effect. Fig.~\ref{Fig:Hall}~(b) is the field dependent transverse resistivity, $\rho_{xy}(H)$ after anti-symmetrization. It shows a steep increase at low magnetic field ($\sim$3~kOe) whereas the slope is reduces in the high field region. The first one is observed due to the anomalous Hall effect (AHE) in the low field regime and the later one, sluggish increase with applied filed, arises because of ordinary Hall effect above 3~kOe. The $\rho_{xy}$ is expressed as
\begin{equation}\label{eq.Hall}
 \rho_{xy}=\rho_{OH}+\rho_{AH}=R_0H+R_sM_s
\end{equation}
where $\rho_{OH}$ is ordinary Hall resistivity and $R_0$ is ordinary Hall coefficient. On the other hand, $\rho_{AH}$, $H$, and $M_{\rm S}$ are anomalous Hall resistivity, applied field and spontaneous magnetization, respectively~\cite{PhysRev.42.709}. Both the contributions, $R_0$ and $\rho_{AH}$, are determined from the slope and intercept of $y$ axis of the linear fit in the high field region. The Hall coefficient, $R_0$ is defined as inverse of carrier concentration ($n$) and electronic charge($e$) $i.e; R_0=1/ne$~\cite{Wang2018}. The positive value of $R_0$ reveals that hole is the dominant charge carrier of the compound. The temperature dependence carrier concentration is shown in Fig.~\ref{Fig:Hall}~(j). The obtained value of $n$ is in the order of 10$^{20}$ cm$^{-3}$. The estimated anomalous Hall resistivity, $\rho_{AH}$ at different temperatures are plotted in Fig.~\ref{Fig:Hall}~(g). The $\rho_{AH}(T)$ follows the spontaneous magnetization curve. It substantiates the ferromagnetism, where the anomalous Hall usually attributed to spontaneous magnetization.

The anomalous Hall effect arises owing to the intrinsic Karplus-Luttinger(KL) and Berry-phase curvature and extrinsic mechanisms such as side-jump and skew scattering. The skew scattering part varies linearly with longitudinal resistivity ($\varpropto\rho_{\rm xx}$), whereas both the intrinsic and the side jump contribution depends quadratically ($\varpropto\rho_{\rm xx}^2$). Thus, $\rho_{AH}$ can be described as
\begin{equation}\label{eq.AHE }
 \rho_{AH}= \sigma^{sk}\rho_{\rm xx}+ \sigma^{sj}\rho_{\rm xx}^2+\sigma^{int}\rho_{\rm xx}^2 = a\rho_{\rm xx}+b\rho_{\rm xx}^2
\end{equation}
where $a$ is the skew scattering and $b$ is the total contribution from the intrinsic and side jump scattering coefficient~\cite{RevModPhys.82.1539}. Due to $\rho_{\rm xx}^2$ dependence of both intrinsic and extrinsic side jump contribution to anomalous Hall value, experimentally it is difficult to differentiate these components. However, to disentangle the skew scattering and the rest of the two contributions, the $\rho_{\rm AH}/\rho_{\rm xx}$ values obtained at different temperatures are plotted against $\rho_{\rm xx}$ (Fig.~\ref{Fig:Hall}~(c))~\cite{PhysRevLett.109.066402,PhysRevB.72.060412,PhysRevB.106.245137}. From linear fitting, the estimated value of $a$ and $b$ is  0.077 and -207.53 $\Omega^{-1}\rm {cm}^{-1}$  respectively. The sign of $a$ and $b$ is opposite, indicating the skew scattering and the combined intrinsic and side jump contributes oppositely to AHE~\cite{adfm.201808747,PhysRevB.110.104407}, while the negetive sign of $b$ can also be a manifestation of atomic disorder of the compound, as found in other disordered HA~\cite{PhysRevB.83.174410}.  By using these coefficients the contribution of skew scattering ($a\rho_{\rm xx}$) and combined contribution of the intrinsic and side-jump ($b\rho_{\rm xx}^2$) is plotted in the same scale (Fig.~\ref{Fig:Hall}~(f)). The graph shows the extrinsic skew scattering dominates over the side jump plus intrinsic contribution in Al$_2$MnFe.

Further,the dominance of the skew scattering term is also investigated in the anomalous Hall conductivity (AHC) behaviour. Accordingly, The Hall conductivity, $\sigma_{yx}$ of the sample has been calculated using the tensor conversion formula~\cite{PhysRevX.8.041045,PhysRevB.107.125138}
\begin{equation}\label{eq.sigma_yx}
 \sigma_{\rm yx}=\frac{\rho_{\rm xy}}{(\rho_{\rm xx}^2+\rho_{\rm xy}^2)}
\end{equation}
The field-dependent Hall conductivity at various temperatures is shown in Fig.~\ref{Fig:Hall}~(e). The anomalous Hall conductivity (AHC), $\sigma_{\rm AH}$ value has been extracted by extrapolating the high field hall conductivity at $y$-axis (Fig.~\ref{Fig:Hall}~(i)). The temperature dependence AHC is plotted in Fig.~\ref{Fig:Hall}~(h). It shows the $\sigma_{\rm AH}$ decreases with temperature and the highest value of AHC for Al$_2$MnFe at 5~K is estimated as $\sim$ 69 S/cm. To understand the relative strength of anomalous Hall current with respect to the normal current, the anomalous Hall angle, $\Theta_{\rm AH}$ = $\sigma_{\rm AH} / \sigma_{\rm xx}$~\cite{PhysRevB.107.125138}, has been calculated, which yielding a maximum value of 1.93\% at 5~K. However, the linear dependence of $\sigma_{\rm AH}$ with respect to $\sigma_{\rm xx}$ (Fig.~\ref{Fig:Hall}~(k)) further establishes the dominance of skew scattering contribution in AHC~\cite{PhysRevB.111.125112}.

\subsection{Heat capacity}
\label{Heat capacity}
The metallic character of a system arise from its density of states (DOS) near the Fermi energy. Heat capacity measurement can provide valuable insight about the DOS. Fig.~\ref{Fig:Mossbauer-resistivity}~(i) presents the heat capacity results of Al$_2$MnFe in the temperature range \SIrange{2}{300}{K}.  The data attains the value of 3$n$R $\sim$ \SI{100}{J.mol^{-1}.K^{-1}} at 300~K which is consistent with the Dulong-Petit law and can be described well by a combination of Debye and Einstein models involving phononic contribution~\cite{PhysRevB.106.224427}. However, to get a more accurate value of electronic contribution of specific heat, we have concentrated on the low temperature region, where the data was fitted by a simple formule, $\gamma T+ \beta T^3$, where $\gamma$ is the electronic contribution and $\beta$ is the phononic part~\cite{Samanta2022}. Our fit yields the value of $\gamma$ and $\beta$ to be   \SI{51.51}{mJ.mol^{-1}.K^{-2}} and \SI{0.146}{mJ.mol^{-1}.K^{-4}}, respectively. A similar value of $\gamma$ has earlier been reported in many other metallic Heusler systems, establishing the bulk metallic character for Al$_2$MnFe as well~\cite{Samanta2022}. Further, a peak is observed in $C_{\rm P}/T$ vs. $T$ plot at $T_{\rm C}$, corroborating magnetic measurements (inset of Fig.~\ref{Fig:Mossbauer-resistivity}~(i)).

\section{Theoretical Results}
\label{Theory}
\begin{table}[b]
\centering
\renewcommand{\arraystretch}{1.0}
\setlength{\tabcolsep}{2pt}
\small
\resizebox{\columnwidth}{!}{
\begin{tabular}{|c|c|c|c|c|c|c|c|}

\hline \hline

& \multicolumn{2}{c|}{$\rule{0pt}{3.0ex}\boldsymbol{M (4a)}$} & \multicolumn{2}{c|}{$\boldsymbol{M (4b)}$} & $\boldsymbol{M_{t}}$
& $\boldsymbol{E_{\mathrm{P}} - E_{\mathrm{AP}}}$
& $\rule{0pt}{2.2ex}\boldsymbol{E_{\mathrm{B2}}^{\mathrm{P}} - E_{\mathrm{L2_1}}^{\mathrm{P}}}$ \\

    & \multicolumn{2}{c|}{\scriptsize{($\mu_{B}$)}}  & \multicolumn{2}{c|}{\scriptsize{($\mu_{B}$)}}  & \scriptsize{($\mu_{B}/f.u.$)}  & \scriptsize{(meV/atom)} & \scriptsize{(meV/atom)}  \\ \hline
\rule{0pt}{1.8ex}\textbf{L2$_1$} & \multicolumn{2}{c|}{\textbf{Mn}}           & \multicolumn{2}{c|}{\textbf{Fe}}           &                          &   & \\ \hline

\textbf{P}  (\scriptsize{$4a\uparrow$ - $4b\uparrow$})             & \multicolumn{2}{c|}{2.46}    & \multicolumn{2}{c|}{0.36}   & 2.78          &  \textbf{-15.82} &   \\ \hline
AP  (\scriptsize{$4a\downarrow$ - $4b\uparrow$})                   & \multicolumn{2}{c|}{-2.28}   & \multicolumn{2}{c|}{0.01}   & -2.25         & 0.0 &  \\ \hline \hline

\rule{0pt}{2.0ex}\textbf{B2} &  \textbf{Mn}         & \textbf{Fe}     & \textbf{Mn}     & \textbf{Fe}    &        &   & \\ \hline
\textbf{P} (\scriptsize{$4a\uparrow$ - $4b\uparrow$})      &  2.38       & 0.34   & 2.40   & 0.31  & 2.69      &   \textbf{-12.78} & \textbf{-18.74} \\ \hline
AP (\scriptsize{$4a\downarrow$ - $4b\uparrow$})            & -2.28       & 0.72   & 2.25   & -0.62 & 0.03      &   0.0 &  \\ \hline \hline
\end{tabular}}

\vspace{1.2ex}  

\resizebox{\columnwidth}{!}{
\begin{tabular}{|c|c|c|c|c|c|c|c|}
\hline \hline
    & \multicolumn{2}{c|}{$\rule{0pt}{2.5ex}\boldsymbol{M (4a)}$}     & \multicolumn{2}{c|}{$\boldsymbol{M (4b)}$}     & \multicolumn{1}{c|}{$\boldsymbol{M (8c)}$}     & $\boldsymbol{M_{t}}$ &  $\boldsymbol{E_{\mathrm{P}}-E_{\mathrm{AP}}}$ \\
    & \multicolumn{2}{c|}{\scriptsize{($\mu_{B}$)}}  & \multicolumn{2}{c|}{\scriptsize{($\mu_{B}$)}}  & \multicolumn{1}{c|}{\scriptsize{($\mu_{B}$)}}  & \scriptsize{($\mu_{B}/f.u.$)}  & \scriptsize{(meV/atom)}  \\ \hline
  \rule{0pt}{2.0ex}\textbf{B2 +12\% Mn-Al}     & \textbf{Mn}   & \textbf{Fe}    & \textbf{Mn}    & \textbf{Fe}    &  \textbf{Mn}      &    &  \\ \hline
P \scriptsize{(($4a,4b)\uparrow$ - $8c\uparrow$})                             & 2.15 & 0.02  & 2.16  & 0.01  &   2.00   &  2.11              &  11.92\\ \hline
\textbf{AP} \scriptsize{(($4a,4b)\uparrow$ - $8c\downarrow$})                 & 2.22 & 0.09 & 2.20  & 0.11 &  -2.61   &  \textbf{1.16}     &  \textbf{0.0} \\ \hline \hline
\end{tabular}}

\caption{Calculated total energy differences  ($E_{\mathrm{P}} - E_{\mathrm{AP}}$) between the parallel (P) and antiparallel (AP) alignments of Mn and Fe atomic moments for the L2$_1$ phase, and among Mn atoms for the B2 phases. The orientation of Fe atomic moments is not taken into account for B2 since their contribution is negligible compared to Mn atoms at different lattice sites. A zero or negative value in the energy difference indicates the relative stability of that phase. The corresponding total magnetic moment (\(M_t\)) and average  atomic magnetic moments are also provided.}
\label{tab:theory}
\end{table}

\begin{figure}[h]
  \centering
  \includegraphics[width=0.9\columnwidth]{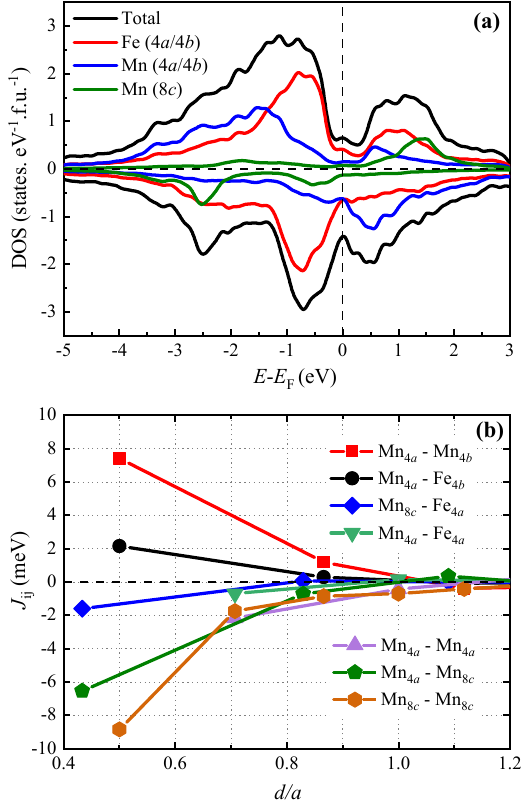}
  \caption{(a) Total and average atom-projected density of states (DOS) for Al$_{2}$FeMn with B2-type along with 12.5\% Al-Mn disorder. The zero of the energy is set at Fermi energy (E$_{F}$). (b) Magnetic exchange interactions ($J_{ij}$) as a function of interatomic distance normalized by the lattice parameter ($d/a$).}\label{Fig:dos}
\end{figure}

Due to the similar X-ray scattering factors of Mn and Fe, the XRD analysis resulted in two ambiguities: (i)  distinguishing between ordered L2$_1$ and B2-type disordered structures, and (ii) determining whether the additional 12\% disorder involves Al swapping with Mn or Fe. To resolve these ambiguities and gain a deeper understanding of the magnetic behavior, we performed first-principles calculations based on density functional theory (DFT).

To address the first issue, we analyzed the energy with different magnetic configurations for both L2$_1$ ordered and B2-type disordered structures and the corresponding results are presented in Table~\ref{tab:theory}. In the L2$_1$ structure, Mn at the 4$a$ sites and Fe at the 4$b$ sites are ferromagnetically coupled. In the B2-type disorder, Mn atoms occupy both the 4$a$ and 4$b$ Wyckoff positions (octahedral sites), with their atomic moments aligning either parallel or anti-parallel. Our results indicate that the most stable configuration corresponds to a parallel alignment of Mn moments at both 4$a$ and 4$b$ sites, along with parallel Fe moments. Furthermore, energy comparisons between the L2$_1$ and B2-type disordered structure reveal that the B2-type disordered structure is energetically more favorable than the L2$_1$ structure by 18.74 meV/atom (Table~\ref{tab:theory}). Thus, energy minimization in Al$_2$MnFe establishes B2-type disordered structure, aligning with the previously reported B2 structure for a closely related composition, Al$_{50}$Mn$_{30}$Fe$_{20}$~\cite{adfm.202310047}.

To investigate the 12\% Al disorder,  whether with Mn or Fe, both disorders are investigated with B2-type structure (Table~\ref{tab:theory}). However, due to the limitations of the supercell size, the disorder was set to 12.5\%, which is in close agreement with the experimental estimation. In this disorder, 12.5\% of either Fe or Mn occupies the Al position at 8$c$, while an equivalent amount of Al replaces Fe or Mn at both the 4$a$ and 4$b$ sites. As observed in B2-type disorder, Mn and Fe favor ferromagnetic coupling. Here, with the additional 12.5\% disorder, the atomic moment of Fe or Mn at the tetrahedral (8$c$) site can be either parallel or anti-parallel to the Fe or Mn atoms at the octahedral sites. In the case of Al-Fe disorder, we see that the Fe at the 8$c$ site exhibits a substantial magnetic moment ($\sim$2.2 $\mu_B$), which should be detectable as a hyperfine field in Mössbauer spectroscopy measurements. However, the absence of any sextet in experimentally observed M\"ossbauer spectra rules out the Fe moment and, consequently, the Al-Fe disorder. In the case of Al-Mn disorder, no significant moment was obtained associated with Fe, corroborating the M\"ossbauer result. Further, we found that the Mn at the 8$c$ site favours an anti-parallel alignment with the Mn at the 4$a$ or 4$b$ sites. Therefore, the Mn atoms at the 4$a$ and 4$b$ sites prefer ferromagnetic coupling, while they prefer ferrimagnetic coupling with the Mn at the 8$c$ site. This is consistent with previous observations studied in similar scenarios~\cite{MaPRB11, KunduPRB17}.  However, the calculated total magnetic moment of 1.16 $\mu_{\mathrm B}$ is slightly lower than what we observed in experiments (1.7 $\mu_{\mathrm B}$). This discrepancy may arise from the specific sampling of Al-Mn disorder, which does not perfectly match the experimental conditions. A statistical average over multiple disordered magnetic configurations, along with an accurate estimation of the disorder concentration, could better reproduce the experimentally observed total magnetic moment. Moreover, the presence of a glassy phase, as identified in experiments, complicates the precise modeling of spin structure. This is because theoretical approaches generally assume a unique ground state spin configuration, whereas spin-glass systems exhibit a degenerate ground state, making accurate comparisons challenging.

Our theoretical investigation shows that Al$_2$MnFe crystallizes in a B2-type disordered structure with 12.5\% Al-Mn disorder. In the ordered L2$_1$ structure, magnetic moment aligns ferromagnetically. Even with B2-type disorder on octahedral sites (4$a$ and 4$b$), the alignment of Mn moment remains ferromagnetic. However, a small 12.5\% disorder between octahedral (4$a$ and 4$b$) and tetrahedral (8$c$) sites induces ferrimagnetic coupling between Mn atoms at 8$c$ and the Mn atoms at 4$a$ and 4$b$. To further understand the microscopic origin of the obtained magnetic moments, the total and atom-projected density of states (DOS) for the B2-type disordered structure with 12.5\% Al-Mn disorder is presented in Fig.~\ref{Fig:dos}~(a). The compensating spin-up and spin-down states describing the low atomic moment of Fe, consistent with M\"ossbauer spectroscopy results. Mn atoms at 8$c$ exhibit a higher atomic moment than those at 4$a$ or 4$b$. This can be attributed to the more prominent exchange splitting of Mn at the 8$c$ site compared to Mn at the 4$a$ or 4$b$ site, which occurs due to the strong hybridization between Mn and Fe at the 4$a$ or 4$b$ sites.

The experimentally observed spin-glass state at low temperatures is often linked to competing magnetic exchange interactions. This motivated us to investigate the interatomic and intra-atomic exchange interactions (J${ij}$) in Al$_2$MnFe. Figure~\ref{Fig:dos}~(b) presents J${ij}$ values (up to the 7$^{\rm th}$ nearest neighbor) as a function of interatomic distance. Notably, Mn atoms dominate the magnetism in Al$_2$MnFe, with most occupying the octahedral 4$a$ and 4$b$ sites. Consequently, the primary exchange coupling in this system arises from the Mn$_{4a}$-Mn$_{4b}$ interaction. The positive value of this interaction plays a crucial role in establishing long-range ferromagnetic ordering in the system. On the other hand, a small fraction (~12\%) of Mn atoms occupy the tetrahedral 8$c$ site. Interestingly, all the exchange interactions involving the Mn atoms at the tetrahedral site are negative, indicating antiferromagnetic (AFM) coupling. The key contributors to this AFM exchange are the exchange coupling between Mn$_{8c}$-Mn$_{4a}$ and Mn$_{8c}$-Mn$_{8c}$, which introduce competing exchange effects within the system. As temperature decreases and thermal fluctuations diminish, these competing interactions become more pronounced. Also, due to disorder, Mn atoms at the 8$c$ sites are randomly distributed. As a result, the long-range FM ordering mediated by octahedrally coordinated Mn$_{4a}$-Mn$_{4b}$ atoms becomes spatially constrained, leading to the emergence of a short-range ordered glassy state at the phase boundary involving Mn atoms at the tetrahedral site.

The role of atomic disorder involving magnetic Mn atoms in the inter-symmetry sites on the evolution of magnetically glassy phase at low temperature can further be supported by comparing the properties of another anti-Heusler compound, Al$_2$MnCu, that remains ferromagnetic down to the lowest temperature~\cite{PhysRevB.111.174417}. Both the compounds exhibit similar structural disorder: B2 + 12\% Mn-Al in the former case, and B2 + 5\% Cu-Al in the latter and Mn is the only magnetic moment-carrying atom in both the materials. In Al$_2$MnCu, the said disorder occurs between the non-magnetic Cu and Al atoms; hence, the basic nature of the magnetic state remains unaltered, and no spin-glass behavior is observed. Contrastingly, in the studied compound Al$_2$MnFe, such disorder involves the magnetic atom Mn, leading to the emergence of a magnetic glassy phase. Thus, the inter-symmetry site disorder, that too involving with moment-carrying atom, is crucial to destabilise the ferromagnetic state in the system.

\section{Discussion and Conclusion}
In summary, this study presents a thorough investigation of the structural, magnetic, and transport properties of the anti-Heusler compound Al$_2$MnFe, highlighting the critical influence of atomic disorder on its physical behavior. The material is confirmed to crystalize in a single-phase cubic Heusler structure, exhibiting two distinct types of disorder: B2-type disorder within octahedral sites and a smaller, yet having a significant impact, Al-Mn disorder between octahedral and tetrahedral sites. Different magnetic measurements firmly establish that the Mn-atoms in this compound undergoes a ferromagnetic transition below $T_{\rm C} \sim$113~K, followed by a development of a reentrant spin-glass phase below $T_{\rm f} \sim$20~K.



This twofold magnetic phase transitions are manifested in transport properties too. The para- to ferromagnetic transition is reflected as a distinct change in slope near the Curie temperature $T_{\rm C}$ on the metallic nature of longitudinal resistivity. Additionally, $\sim{H^{\rm2/3}}$ dependence of magnetoresistance around $T_{\rm C}$ also points toward ferromagnetic interactions and $s–d$ scattering of conduction electrons. Furthermore, the compound exhibits considerable anomalous Hall conductivity, 69 S/cm, providing additional evidence of the material's underlying ferromagnetic nature. A clear dip in the d$\rho_{xy}$/d$T$ precisely identifies the $T_{\rm C}$, corroborating the magnetic measurements, whereas the thermal irreversibility of the temperature dependent ZFC and FC transverse Hall resistivity curves supports the appearance of the glassy magnetic state at low temperatures.

Theoretical analysis reveals a very fascinating feature by exposing a rather limited influence of atomic disorder of moment carrying Mn atoms on the ferromagnetic interaction, when the disorder is confined within the octahedral sites ($4a$ and $4b$), even when the disorder is maximum. On the other hand, a relatively small ($\sim$12\%) transfer of Mn atoms to the tetrahedral (8$c$) sites appears sufficient to introduce antiferromagnetic interactions, resulting in a reentrant spin-glass phase at low temperatures.


Our findings pave the way for further exploration of anti-Heusler alloys, where controlled disorder could be strategically harnessed to design materials with tailored magnetic and electronic properties, with promising implications for next-generation spintronic applications. This dual-phase magnetic behavior not only deepens the understanding of reentrant spin-glass states in disordered systems but also offers valuable insights for tuning magnetic interactions via atomic site engineering.

\begin{acknowledgments}
Soumya Bhowmik would like to sincerely acknowledge SINP, India for the fellowship and research facility. The authors acknowledge A. Thamizhavel and R. Kulkarni for heat capacity measurement and S. Dan for fruitful discussions. A. K. and M. K. gratefully acknowledge the computational resources provided by the PARAM Brahma Facility at IISER Pune, under the National Supercomputing Mission of the Government of India.
\end{acknowledgments}

\normalem
\bibliographystyle{apsrev4-2}
%

\end{document}